\documentclass[twocolumn,nofootinbib,superscriptaddress,floatfix]{revtex4-1}

\usepackage{array}

\usepackage[totalwidth=480pt, totalheight=680pt]{geometry}
\usepackage{setspace}

\usepackage{amsmath}
\usepackage{mathtools}
\usepackage{epstopdf}

\usepackage{verbatim}
\usepackage{booktabs}

\usepackage[usenames,dvipsnames]{color}
\usepackage{color}
\usepackage{pstricks,framed}

\usepackage{tensor}
\usepackage{amssymb}
\usepackage{amsthm}
\usepackage{amsfonts}
\usepackage{simplewick}

\usepackage{chngcntr}
\usepackage[retainorgcmds]{IEEEtrantools}
\usepackage{caption}

\usepackage{graphicx}
\usepackage[colorlinks=true, linktoc=page, citecolor=blue, urlcolor=blue]{hyperref}

\counterwithin{equation}{section}
\makeatletter
\def\p@subsection{}
\def\p@subsubsection{}
\makeatother

\allowdisplaybreaks

\begin{document}

\title{\color{Blue}\textbf{Multipole stability of spinning M2-branes in the \\ classical limit of the BMN matrix model}}

\author{\textbf{Minos Axenides}}
\email{axenides@inp.demokritos.gr}
\affiliation{Institute of Nuclear and Particle Physics, N.C.S.R.\ ``Demokritos",\\ 153 10, Agia Paraskevi, Greece}
\author{\textbf{Emmanuel Floratos}}
\email{mflorato@phys.uoa.gr}
\affiliation{Institute of Nuclear and Particle Physics, N.C.S.R.\ ``Demokritos",\\ 153 10, Agia Paraskevi, Greece}
\affiliation{Department of Physics, National and Kapodistrian University of Athens,\\ Zografou Campus, 157 84, Athens, Greece}
\author{\textbf{Georgios Linardopoulos}}
\email{glinard@inp.demokritos.gr}
\affiliation{Institute of Nuclear and Particle Physics, N.C.S.R.\ ``Demokritos",\\ 153 10, Agia Paraskevi, Greece}
\affiliation{Department of Physics, National and Kapodistrian University of Athens,\\ Zografou Campus, 157 84, Athens, Greece}

\begin{abstract}
\normalsize{\noindent We explore the stability of a recently found class of spinning dielectric M2-branes in the 11-dimensional maximally supersymmetric plane-wave background. We find two small windows of instabilities in the dipole ($j=1$) and quadrupole ($j = 2$) sector of linear multipole perturbations.}
\end{abstract}

\maketitle
\section[Introduction]{Introduction}
\noindent There is strong evidence \cite{SekinoSusskind08} that matrix theory \cite{BFSS97, BMN02} provides a valid description of the chaotic and non-local dynamics of the microscopic degrees of freedom that are present on the horizons of black holes. Superfast propagation and mixing of information (fast scrambling) \cite{HaydenPreskill07, SekinoSusskind08} are expected to be emergent features of quantum matrix models that are known to reduce to M2-branes in the continuum limit \cite{deWitHoppeNicolai88, DasguptaJabbariRaamsdonk02}. \\[6pt]
\indent In a recent letter \cite{AxenidesFloratosLinardopoulos17a}, we identified and studied a class of ellipsoidal membranes that spin inside the 11-dimensional maximally supersymmetric plane-wave background \cite{KowalskiGlikman84a, OFarrillPapadopoulos02b}:
\begin{IEEEeqnarray}{l}
ds^2 = -2 dx^{+} dx^{-} + \sum_{i=1}^3 dx_i dx_i + \sum_{k=1}^6 dy_k dy_k - \nonumber \\
\hspace{1cm}- \left[\frac{\mu^2}{9}\sum_{i=1}^3 x_i x_i + \frac{\mu^2}{36}\sum_{k=1}^6 y_k y_k\right] dx^+ dx^+ \qquad \label{MaximallySupersymmetricMetric} \\
F_{123+} = \mu. \label{MaximallySupersymmetricMetricFieldStrength}
\end{IEEEeqnarray}
The equations of motion of a membrane in the above background read, in the light-cone gauge ($x^+ \equiv \tau$):
\begin{IEEEeqnarray}{ll}
\ddot{x}_i = &\left\{\left\{x_i,x_k\right\},x_k\right\} + \left\{\left\{x_i,y_k\right\},y_k\right\} - \frac{\mu^2}{9}\,x_i + \nonumber \\
& +\frac{\mu}{2}\epsilon_{ikl}\left\{x_k,x_l\right\}, \quad i = 1,2,3 \label{xEquation} \\
\ddot{y}_i = &\left\{\left\{y_i,y_k\right\},y_k\right\} + \left\{\left\{y_i,x_k\right\},x_k\right\} - \frac{\mu^2}{36}\,y_i, \qquad \nonumber \\
& i = 1,\ldots,6. \label{yEquation}
\end{IEEEeqnarray}
The corresponding Gauss-law constraint is given by
\begin{IEEEeqnarray}{l}
\left\{\dot{x}_i, x_i\right\} + \left\{\dot{y}_k, y_k\right\} = 0. \label{GaussLaw1}
\end{IEEEeqnarray}
\indent A large number of solutions of the BMN matrix model is known at both finite and large values of the matrix dimensionality $N$. See \cite{Bak02a, Mikhailov02b, Park02b, BakKimLee05, HoppeLee07} for BPS solutions of various topologies and \cite{ArnlindHoppe03b, ArnlindHoppeTheisen04, BerensteinDzienkowskiLashof-Regas15, Hoppe15} for other rotating (non-BPS) solutions. \\[6pt]
\indent The membrane solutions that we proposed in \cite{AxenidesFloratosLinardopoulos17a} are very simple. The first one lives exclusively in $SO(3)$:
\begin{IEEEeqnarray}{c}
x_i = \mu u_i e_i, \ i = 1,2,3, \quad y_k = 0, \ k = 1,\ldots 6, \qquad \label{Ansatz1}
\end{IEEEeqnarray}
while the second one has nonzero components in both counterparts of $SO(3)\times SO(6)$:\footnote{This solution is similar to the one that was introduced in \cite{HarmarkSavvidy00} in the context of D0-brane matrix mechanics. In turn, the latter was largely inspired by \cite{CollinsTucker76}.}
\begin{IEEEeqnarray}{ll}
x_i = \tilde{u}_i\left(\tau\right) e_i, \quad & y_k = \tilde{v}_k\left(\tau\right) e_k, \label{Ansatz2} \\[6pt]
& y_{k+3} = \tilde{w}_k\left(\tau\right) e_k, \ i,k = 1,2,3, \qquad \quad \label{Ansatz3}
\end{IEEEeqnarray}
where
\begin{IEEEeqnarray}{ll}
\tilde{u}_i = \mu u\left(t\right), \quad &\tilde{v}_k = \mu v\left(t\right)\cos\left(\omega t + \varphi_k\right) \label{Ansatz4} \\
& \tilde{w}_k = \mu v\left(t\right)\sin\left(\omega t + \varphi_k\right) \qquad \label{Ansatz5}
\end{IEEEeqnarray}
and the three coordinates of the unit sphere,
\begin{IEEEeqnarray}{c}
(e_1, e_2, e_3) = (\cos\phi \sin\theta, \sin\phi \sin\theta, \cos\theta) \nonumber \\
\phi \in [0,2\pi), \quad \theta \in [0,\pi] \label{Epsilon1}
\end{IEEEeqnarray}
satisfy the $\mathfrak{so}\left(3\right)$ Poisson bracket algebra and are orthonormal:
\begin{IEEEeqnarray}{c}
\{e_a, e_b\} = \epsilon_{abc} \, e_c, \quad \int e_a \, e_b\,d^2\sigma = \frac{4\pi}{3} \, \delta_{ab}. \qquad \label{Epsilon2}
\end{IEEEeqnarray}
\indent The effective potential that arises from the $SO(3)$ ansatz \eqref{Ansatz1} is the generalized 3d H\'{e}non-Heiles potential \cite{EfstathiouSadovskii04}. It has exactly nine critical points:
\begin{IEEEeqnarray}{c}
\textbf{u}_0 = 0, \ \textbf{u}_{1/6} = \frac{1}{6}\,\left(1, 1, 1\right), \ \textbf{u}_{1/3} = \frac{1}{3}\,\left(1, 1, 1\right), \qquad \quad \label{So3Extrema}
\end{IEEEeqnarray}
where the remaining six extrema can be obtained from $\textbf{u}_{1/6}$ and $\textbf{u}_{1/3}$ by flipping the signs of exactly two of their coordinates. The discrete symmetry group of the nine extrema is the tetrahedral group $T_d$. This discrete symmetry can be extended to the full equations of motion \eqref{xEquation}--\eqref{GaussLaw1} by applying separate reflections to each coordinate $x_i \rightarrow \epsilon_i x_i$, where $\epsilon_i = \pm 1$. The flux term leads to the constraint $\epsilon_1\epsilon_2\epsilon_3 = 1$. \\[6pt]
\indent Now we showed in \cite{AxenidesFloratosLinardopoulos17a} that $\textbf{u}_0$ (the collapsed point-like membrane) and $\textbf{u}_{1/3}$ (the Myers dielectric sphere) are radially stable local minima, whereas $\textbf{u}_{1/6}$ is a radially unstable saddle point. \\[6pt]
\indent For $u \equiv u_1 = u_2 = u_3$, \eqref{Ansatz1} leads to a double-well potential. The corresponding solutions can be expressed in terms of the Jacobian elliptic functions:
\begin{IEEEeqnarray}{ll}
u\left(t\right) = &\frac{1}{6} \pm \sqrt{\frac{1}{6^2} + \sqrt{\mathcal{E}}} \cdot \nonumber \\
&\cdot cn\left[\sqrt{2\sqrt{\mathcal{E}}}\cdot t\Bigg|\frac{1}{2}\left(1 + \frac{1}{36\sqrt{\mathcal{E}}}\right)\right], \qquad \ \label{OrbitSO3}
\end{IEEEeqnarray}
where $\mathcal{E} \equiv E/(2\pi T\mu^4)$ and $T$ is the brane tension. \\[6pt]
\indent On the other hand, the effective potential that arises from \eqref{Ansatz2}--\eqref{Ansatz5} always possesses a continuous set of critical points $(u_0,v_0)$ inside the interval:
\begin{IEEEeqnarray}{ll}
\frac{1}{6} \leq u_0 \leq \frac{1}{3} \qquad \& \qquad 0 \leq v_0 \leq \frac{1}{12}, \qquad \label{ExtremalBounds}
\end{IEEEeqnarray}
where
\begin{IEEEeqnarray}{l}
v_0^2 = \left(u_0 - \frac{1}{6}\right)\left(\frac{1}{3} - u_0\right), \quad \omega^2 = u_0 - \frac{1}{12}. \qquad \quad \label{ExtremalPoints}
\end{IEEEeqnarray}
It can be shown that the energy and the angular momentum ($\ell^2 \equiv \omega v_0^2$) of the solutions \eqref{Ansatz2}--\eqref{Ansatz5} are bounded from above by their values at $u_{\text{crit}}$. The extrema $(u_0,v_0)$ are radially stable inside the interval:
\begin{IEEEeqnarray}{c}
u_{\text{crit}} \leq u_0 \leq \frac{1}{3}, \qquad u_{\text{crit}} = \frac{1}{60} \left(11+\sqrt{21}\right) \qquad \label{StabilityInterval}
\end{IEEEeqnarray}
and (radially) unstable outside of it. \\[6pt]
\indent The purpose of the present letter is to examine the stability of the two configurations \eqref{Ansatz1} and \eqref{Ansatz2}--\eqref{Ansatz5} under linearized multipole perturbations. This is carried out in the following section \ref{Section:AngularPerturbations}. For simplicity we focus exclusively on the bosonic sector. Section \ref{Section:Discussion} contains a brief discussion of our results.
\section[Stability Analysis]{Stability Analysis \label{Section:AngularPerturbations}}
\subsection[$SO(3)$ Sector]{$SO(3)$ Sector \label{SubSection:AngularPerturbations1}}
\noindent As a warm-up, we examine the multipole stability of the $SO(3)$ solution \eqref{Ansatz1}. Our analysis parallels the one for the BMN matrix model in \cite{DasguptaJabbariRaamsdonk02}, but our perspective is completely different. For the most part we will be following a method that was introduced in \cite{AxenidesFloratosPerivolaropoulos00, AxenidesFloratosPerivolaropoulos01} for the stability of membranes in flat backgrounds.
\newpage
\indent Consider the following set of linearized perturbations around the classical solution \eqref{Ansatz1}:
\begin{IEEEeqnarray}{ll}
x_i = x_i^0 + \delta x_i, \quad &i = 1, 2, 3 \label{AngularPerturbations1a} \\
y_k = \delta y_k, \quad &k = 1,\ldots 6, \label{AngularPerturbations1b}
\end{IEEEeqnarray}
where the (static) solutions $x_i^0 = \mu u_0 e_i$ ($i = 1,2,3$) are constructed from the spherically symmetric extrema $u_0 \equiv u_1^0 = u_2^0 = u_3^0$ in \eqref{So3Extrema}. Plugging \eqref{AngularPerturbations1a}--\eqref{AngularPerturbations1b} into the equations of motion \eqref{xEquation}--\eqref{yEquation} we get:
\begin{IEEEeqnarray}{l}
\delta\ddot{x}_i = \left\{\left\{\delta x_i,x_k^0\right\},x_k^0\right\} + \left\{\left\{x_i^0,\delta x_k\right\},x_k^0\right\} + \nonumber \label{PerturbationEquation1} \\
+ \left\{\left\{x_i^0,x_k^0\right\},\delta x_k\right\} - \frac{\mu^2}{9}\,\delta x_i + \mu\epsilon_{ikl}\left\{\delta x_k,x_l^0\right\} \qquad \\
\delta\ddot{y}_i = \left\{\left\{\delta y_i,x_k^0\right\},x_k^0\right\} - \frac{\mu^2}{36}\,\delta y_i, \label{PerturbationEquation2}
\end{IEEEeqnarray}
while the linearized Gauss constraint \eqref{GaussLaw1} becomes:
\begin{IEEEeqnarray}{c}
\left\{\delta\dot{x}_i, x_i^0\right\} = 0, \label{GaussLaw2}
\end{IEEEeqnarray}
since $\dot{x}_i^0 = {y}_i^0 = \dot{y}_i^0 = 0$. It can be proven that if \eqref{GaussLaw2} is satisfied at $\tau = 0$, the perturbation equations \eqref{PerturbationEquation1}--\eqref{PerturbationEquation2} guarantee its validity at all times. We will return to this point at the end of the subsection. \\[6pt]
Next, we expand $\delta x$ and $\delta y$ in spherical harmonics:
\begin{IEEEeqnarray}{ll}
\delta x_i = \mu\cdot\sum_{j = 1}^{\infty}\sum_{m=-j}^{j}\eta_i^{jm}\left(\tau\right) Y_{jm}\left(\theta,\phi\right) \qquad \label{AngularPerturbations2} \\
\delta y_k = \mu\cdot\sum_{j = 1}^{\infty}\sum_{m=-j}^{j}\theta_k^{jm}\left(\tau\right) Y_{jm}\left(\theta,\phi\right), \qquad \label{AngularPerturbations3}
\end{IEEEeqnarray}
where again $i = 1,2,3$ and $k = 1,\ldots,6$. We use the following property of spherical harmonics:
\begin{IEEEeqnarray}{c}
\left\{e_i,Y_{jm}\left(\theta,\phi\right)\right\} = -i \hat{J}_i Y_{jm}\left(\theta,\phi\right), \qquad \label{SphericaHarmonics1}
\end{IEEEeqnarray}
where $\hat{J}_i$ is the angular momentum operator in spherical coordinates. In matrix form \eqref{SphericaHarmonics1} is written as:
\begin{IEEEeqnarray}{c}
\left\{e_i,Y_{jm}\left(\theta,\phi\right)\right\} = -i \sum_{m'}\left(J_i\right)_{m'm} Y_{jm'}\left(\theta,\phi\right), \qquad \label{SphericaHarmonics2}
\end{IEEEeqnarray}
where $\left(J_i\right)_{mm'}$ furnish a $2j+1$ dimensional matrix representation of $\mathfrak{su}\left(2\right)$. Using \eqref{SphericaHarmonics1}--\eqref{SphericaHarmonics2} we can show that the fluctuation modes $\eta_i$ and $\theta_i$ satisfy the following $2j+1$ dimensional equations of motion:
\begin{IEEEeqnarray}{l}
\ddot\eta_i + \omega_3^2 \eta_i = u_0^2 \, T_{ik} \eta_k + u_0 Q_{ik} \eta_k \label{AngularPerturbations4a} \\
\ddot\theta_i + \omega_6^2 \theta_i = 0, \label{AngularPerturbations4b}
\end{IEEEeqnarray}
where we have switched to dimensionless time $t \equiv \mu\tau$, have omitted the indices $j,m$ and have used the following definitions:
\begin{IEEEeqnarray}{l}
\omega_3^2 \equiv u_0^2 j\left(j+1\right) + \frac{1}{9}, \quad \omega_6^2 \equiv u_0^2 j\left(j+1\right) + \frac{1}{36} \qquad \quad \label{Definitions1} \\
T_{ik} \equiv J_i J_k - 2i\epsilon_{ikl}J_l, \quad Q_{ik} \equiv i\epsilon_{ikl}J_l. \qquad \label{Definitions2}
\end{IEEEeqnarray}
\eqref{AngularPerturbations4a}--\eqref{AngularPerturbations4b} can be written in compact form as
\begin{IEEEeqnarray}{l}
\ddot{H} + \left(\omega_3^2 I - u_0^2 T - u_0 Q\right)\cdot H = 0 \label{AngularPerturbations5a} \\
\ddot\Theta + \omega_6^2 \Theta = 0, \qquad \label{AngularPerturbations5b}
\end{IEEEeqnarray}
where $H \equiv \left(\eta_i\right)$, $\Theta \equiv \left(\theta_i\right)$, $I \equiv \left(\delta_{ik}\right)$ $Q \equiv \left(Q_{ik}\right)$ and $T \equiv \left(T_{ik}\right)$. By setting
\begin{IEEEeqnarray}{c}
\left[\begin{array}{c} H \\ \Theta \end{array}\right] = e^{i\lambda t} \left[\begin{array}{c} \boldsymbol{\xi}_1 \\ \boldsymbol{\xi}_2 \end{array}\right], \label{Eigenvalues1}
\end{IEEEeqnarray}
we are led to the following eigenvalue problem:
\begin{IEEEeqnarray}{l}
\left[\left(-\lambda^2 + \omega_3^2\right)I - u_0^2 \, T - u_0 \, Q\right] \cdot \boldsymbol{\xi}_1 = 0 \label{AngularPerturbations6a} \\
\left(-\lambda^2 + \omega_6^2\right) \cdot \boldsymbol{\xi}_2 = 0. \label{AngularPerturbations6b}
\end{IEEEeqnarray}
\indent In order to solve \eqref{AngularPerturbations6a}--\eqref{AngularPerturbations6b} we introduce the projection operators:
\small\begin{IEEEeqnarray}{l}
P \equiv \frac{1}{j\left(j+1\right)} \, J_i J_k \label{AngularEigenvalueProblem1a} \\
R_{\pm} \equiv \frac{1}{2j +1} \Big[\frac{1}{2}\left(2j+1\mp1\right)\left(I - P\right) \pm \left(I - Q\right)\Big], \qquad \label{AngularEigenvalueProblem1b}
\end{IEEEeqnarray}\normalsize
which are orthonormal and form a complete set:
\begin{IEEEeqnarray}{l}
I = P + R_+ + R_-\,. \label{AngularEigenvalueProblem2c}
\end{IEEEeqnarray}
The projection operators $P$, $R_{\pm}$ can be used to express the matrices $T$ and $Q$ as follows:
\begin{IEEEeqnarray}{l}
T = \left[j\left(j+1\right) - 2\right]P + 2jR_+ - 2\left(j+1\right)R_- \qquad \label{AngularEigenvalueProblem2a} \\
Q = P - jR_+ + \left(j+1\right)R_-\,. \label{AngularEigenvalueProblem2b}
\end{IEEEeqnarray}
The eigenvalues $\lambda$ are found by solving \eqref{AngularPerturbations6a}--\eqref{AngularPerturbations6b}:
\begin{IEEEeqnarray}{l}
\left(\omega_3^2 - \lambda^2\right)\boldsymbol{\xi}_1 = \bigg[\left(u_0^2\left[j\left(j+1\right) - 2\right] + u_0\right)P + \nonumber \\
j u_0 \left(2u_0 - 1\right)R_+ - \left(j+1\right)u_0\left(2u_0 - 1\right)R_-\bigg]\boldsymbol{\xi}_1, \qquad \quad
\end{IEEEeqnarray}
which leads to
\begin{IEEEeqnarray}{l}
\lambda_P^2 = 2(u_0 - \frac{1}{3})(u_0 - \frac{1}{6}) \label{Eigenfrequency1} \\
\lambda_+^2 = j\left(j-1\right)u_0^2 + j u_0 + \frac{1}{9} \label{Eigenfrequency2} \\
\lambda_-^2 = \left(j+1\right)\left(j+2\right)u_0^2 - \left(j+1\right)u_0 + \frac{1}{9} \qquad \label{Eigenfrequency3} \\
\lambda_{\theta}^2 = u_0^2 j\left(j+1\right) + \frac{1}{36}. \label{Eigenfrequency4}
\end{IEEEeqnarray}
The multiplicities of equations \eqref{Eigenfrequency1}--\eqref{Eigenfrequency3} are equal to the dimensionalities of the corresponding projectors $P$, $R_+$, $R_-$, i.e.\ $d_P = 2j+1$, $d_+ = 2j+3$, $d_- = 2j-1$. The degeneracy of the decoupled $\theta$-oscillators in \eqref{Eigenfrequency4} is equal to $6\left(2j+1\right)$. In total there are $18\left(2j+1\right)$ eigenvalues. For each of the critical points in \eqref{So3Extrema} ($u_0 = 0,1/6,1/3$) we find:
\begin{IEEEeqnarray}{ll}
\textbf{u}_0: \ &\lambda_P^2 = \lambda_{\pm}^2 = \frac{1}{9}, \quad \lambda_{\theta}^2 = \frac{1}{36} \qquad \label{Eigenfrequency5} \\[6pt]
\textbf{u}_{1/6}: \ &\lambda_P^2 = 0, \quad \lambda_+^2 =\frac{1}{36}\left(j + 1\right)\left(j + 4\right) \qquad \label{Eigenfrequency6} \\
&\lambda_-^2 = \frac{j\left(j-3\right)}{36}, \quad \lambda_{\theta}^2 = \frac{1}{36}\left(j^2 + j + 1\right) \qquad \quad \label{Eigenfrequency7} \\[6pt]
\textbf{u}_{1/3}: \ &\lambda_P^2 = 0, \quad \lambda_+^2 =\frac{1}{36}\left(j + 1\right)^2 \qquad \quad \label{Eigenfrequency8} \\
& \lambda_-^2 = \frac{j^2}{9}, \quad \lambda_{\theta}^2 = \frac{1}{36}\left(2j + 1\right)^2. \qquad \label{Eigenfrequency9}
\end{IEEEeqnarray}
\indent We observe that the critical point $\textbf{u}_0$ (the point-like membrane) is stable, $\textbf{u}_{1/3}$ has a zero mode of degeneracy $2d_P$ in the $P$-sector for $j = 1,2,\ldots$ along with its stable eigenvalues in the remaining sectors. The critical point $\textbf{u}_{1/6}$ has one $2d_P$ degenerate zero mode for every $j$ and one 10-fold degenerate zero mode for $j=3$. It is unstable for $j=1$ (2-fold degenerate) and $j=2$ (6-fold degenerate) in the $R_-$ sector. These instabilities correspond to the top of the double-well potential which is radially a saddle point. \\[6pt]
\indent As we mentioned in the introduction, the tetrahedral symmetry of the $SO(3)$ extrema can be extended to the full Hamiltonian of the membrane. This implies that the tetrahedral images of the critical points $\textbf{u}_{1/3}$ and $\textbf{u}_{1/6}$ will have the same spectra under the multipole perturbations \eqref{AngularPerturbations1a}--\eqref{AngularPerturbations1b}, \eqref{AngularPerturbations2}--\eqref{AngularPerturbations3}. \\[6pt]
\indent To prove the validity of the Gauss law \eqref{GaussLaw2}, we insert $x_i^0 = \mu u_0 e_i$ and $\delta x_i(0)$ from \eqref{AngularPerturbations2} into \eqref{GaussLaw2}, and use the property \eqref{SphericaHarmonics2} to get at $t=0$:
\begin{IEEEeqnarray}{ll}
\sum_{m}\sum_{i=1}^3 \left(J_i\right)_{m'm}\dot{\eta}_{i}^{jm}\left(0\right) = 0. \label{CoplanarityConstraint1}
\end{IEEEeqnarray}
This system always has a solution which can be expressed as a linear combination of the eigenvectors $\boldsymbol{\xi}_1$ and $\boldsymbol{\xi}_2$. Therefore the Gauss-law constraint \eqref{GaussLaw2} will be satisfied at $t = 0$ and, according to what we have said before, it will also be satisfied at all times. \\[6pt]
\indent Our results are in agreement with those of \cite{DasguptaJabbariRaamsdonk02}. However our method has significant advantages which will turn up in the following section, where we examine the multipole stability of the much more complicated $SO(3)\times SO(6)$ solution \eqref{Ansatz2}--\eqref{Ansatz5}.
\subsection[$SO(3)\times SO(6)$ Sector]{$SO(3)\times SO(6)$ Sector \label{SubSection:AngularPerturbations2}}
\begin{figure}[t]
\begin{center}
\includegraphics[scale=0.4]{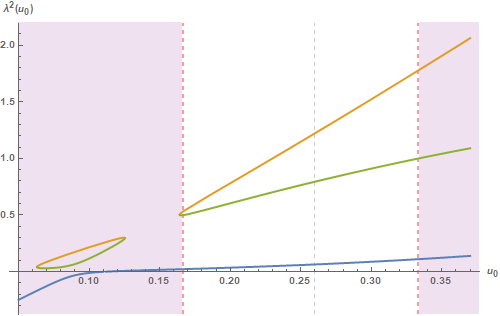}
\caption{Squares of the $j = 2$ eigenvalues of $\mathbb{A}_{+}$ in terms of the coordinate $u_0$.} \label{Graph:Eigenvalues1}
\includegraphics[scale=0.4]{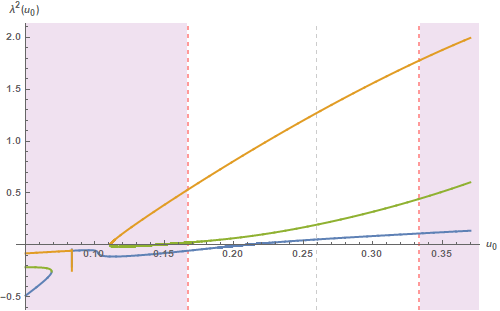}
\caption{Squares of the $j = 2$ eigenvalues of $\mathbb{A}_{-}$ in terms of the coordinate $u_0$.} \label{Graph:Eigenvalues2}
\end{center}
\end{figure}

\noindent To treat the $SO(3)\times SO(6)$ solution \eqref{Ansatz2}--\eqref{Ansatz5}, let us insert the following linear perturbations
\begin{IEEEeqnarray}{ll}
x_i = x_i^0 + \delta x_i, \quad & i = 1, 2, 3 \\
y_i = y_i^0 + \delta y_i, \quad & i = 1,\ldots,6, \label{AngularPerturbations7} \qquad
\end{IEEEeqnarray}
into the equations of motion \eqref{xEquation}--\eqref{yEquation} and the Gauss law \eqref{GaussLaw1}. The fluctuation equations become:
\begin{IEEEeqnarray}{ll}
\delta\ddot{x}_i = &\left\{\left\{\delta x_i,x_j^0\right\},x_j^0\right\} + \left\{\left\{x_i^0,\delta x_j\right\},x_j^0\right\} + \nonumber \\
& + \left\{\left\{x_i^0,x_j^0\right\},\delta x_j\right\} + \left\{\left\{\delta x_i,y_j^0\right\},y_j^0\right\} + \nonumber \\
& + \left\{\left\{x_i^0,\delta y_j\right\},y_j^0\right\} + \left\{\left\{x_i^0,y_j^0\right\},\delta y_j\right\} - \nonumber \\
& - \frac{\mu^2}{9}\,\delta x_i + \mu\epsilon_{ijk}\left\{\delta x_j,x_k^0\right\} \label{PerturbationEquation3} \\[6pt]
\delta\ddot{y}_i = &\left\{\left\{\delta y_i,y_j^0\right\},y_j^0\right\} + \left\{\left\{y_i^0,\delta y_j\right\},y_j^0\right\} + \nonumber \\
& + \left\{\left\{y_i^0,y_j^0\right\},\delta y_j\right\} + \left\{\left\{\delta y_i,x_j^0\right\},x_j^0\right\} + \nonumber \\
& + \left\{\left\{y_i^0,\delta x_j\right\},x_j^0\right\} + \left\{\left\{y_i^0,x_j^0\right\},\delta x_j\right\} - \nonumber \\
& - \frac{\mu^2}{36}\,\delta y_i, \label{PerturbationEquation4}
\end{IEEEeqnarray}
while the Gauss-law constraint reads:
\small\begin{IEEEeqnarray}{c}
\left\{\delta\dot{x}_i, x_i^0\right\} + \left\{\dot{x}_i^0, \delta x_i\right\} + \left\{\delta\dot{y}_i, y_i^0\right\} + \left\{\dot{y}_i^0, \delta y_i\right\} = 0. \qquad \quad \label{GaussLaw3}
\end{IEEEeqnarray} \normalsize
In our case,
\begin{IEEEeqnarray}{ll}
x_i^0 = \mu u_0 e_i, \quad &i = 1, 2, 3 \qquad \\
y_i^0 = \mu v_i^0\left(t\right) e_1, \quad &i = 1,2 \qquad \\
y_k^0 = \mu v_k^0\left(t\right) e_2, \quad &k = 3,4 \qquad \\
y_l^0 = \mu v_l^0\left(t\right) e_3, \quad &l = 5,6, \qquad
\end{IEEEeqnarray}
\newpage where
\begin{IEEEeqnarray}{ll}
v_j^0\left(t\right) = v_0 \cos\left(\omega t + \varphi_j\right), \quad & j = 1,3,5 \qquad \label{DielectricTop1} \\[6pt]
v_k^0\left(t\right) = v_0 \sin\left(\omega t + \varphi_k\right), \quad & k = 2,4,6 \qquad \label{DielectricTop2}
\end{IEEEeqnarray}
and $(u_0,v_0)$ are the extrema of the corresponding effective potential that satisfy \eqref{ExtremalBounds}--\eqref{ExtremalPoints}. Let us again expand $\delta x$ and $\delta y$ in spherical harmonics:
\begin{IEEEeqnarray}{lll}
\delta x_i = \mu\cdot\sum_{j,m}\eta_i^{jm}\left(\tau\right) Y_{jm}\left(\theta,\phi\right) \qquad \label{AngularPerturbations8} \\
\delta y_k = \mu\cdot\sum_{j,m}\epsilon_k^{jm}\left(\tau\right) Y_{jm}\left(\theta,\phi\right) \qquad \label{AngularPerturbations9} \\
\delta y_l = \mu\cdot\sum_{j,m}\zeta_l^{jm}\left(\tau\right) Y_{jm}\left(\theta,\phi\right), \qquad \label{AngularPerturbations10}
\end{IEEEeqnarray}
where $i = 1,2,3$, $k = 1,3,5$ and $l = 2,4,6$. Switching to dimensionless time $t \equiv \mu\tau$ and using \eqref{SphericaHarmonics1}--\eqref{SphericaHarmonics2}, we find the following equations of motion for the fluctuation modes $\eta_i$, $\epsilon_i$ and $\zeta_i$ (omitting the indices $j,m$):
\begin{widetext}
\vspace{-.6cm}\begin{IEEEeqnarray}{l}
\ddot\eta_i + \omega_3^2 \eta_i = u_0 T_{ik}\left(u_0\eta_k + v_0\cos\left(\omega t + \varphi_k\right)\epsilon_k + v_0\sin\left(\omega t + \varphi_k\right)\zeta_k\right) + u_0 Q_{ik} \eta_k \label{AngularPerturbations11} \\[6pt]
\ddot\epsilon_i + \omega_6^2 \epsilon_i = v_0\cos\left(\omega t + \varphi_i\right) T_{ik}\left(u_0\eta_k + v_0\cos\left(\omega t + \varphi_k\right)\epsilon_k + v_0\sin\left(\omega t + \varphi_k\right)\zeta_k\right) \label{AngularPerturbations12} \\[6pt]
\ddot\zeta_i + \omega_6^2 \zeta_i = v_0\sin\left(\omega t + \varphi_i\right) T_{ik}\left(u_0\eta_k + v_0\cos\left(\omega t + \varphi_k\right)\epsilon_k + v_0\sin\left(\omega t + \varphi_k\right)\zeta_k\right), \label{AngularPerturbations13}
\end{IEEEeqnarray}
\end{widetext}
where $T_{ik}$ and $Q_{ik}$ have been defined in \eqref{Definitions2} and summation is implied over all the repeated indices except $i$. We have also used the definitions:
\begin{IEEEeqnarray}{l}
\omega_3^2 \equiv \left(u_0^2 + v_0^2\right) j\left(j+1\right) + \frac{1}{9} \quad \label{Definitions3} \\
\omega_6^2 \equiv \left(u_0^2 + v_0^2\right) j\left(j+1\right) + \frac{1}{36}. \qquad \label{Definitions4}
\end{IEEEeqnarray}
We can transform \eqref{AngularPerturbations11}--\eqref{AngularPerturbations13} into a set of equations with constant coefficients by making the following rotation in the $\left(\epsilon_i,\zeta_i\right)$ space:
\begin{IEEEeqnarray}{l}
\theta_i = \epsilon_i \cdot \cos\left(\omega t + \varphi_i\right) + \zeta_i \cdot \sin\left(\omega t + \varphi_i\right) \qquad \label{Rotation1} \\
\chi_i = -\epsilon_i \cdot \sin\left(\omega t + \varphi_i\right) + \zeta_i \cdot \cos\left(\omega t + \varphi_i\right). \qquad \label{Rotation2}
\end{IEEEeqnarray}
We find:
\small\begin{IEEEeqnarray}{l}
\ddot{H} + \left(\omega_3^2 \, I - u_0^2 \, T - u_0 \, Q\right) \cdot H - u_0 \, v_0 \, T\cdot\Theta = 0 \qquad \\
\ddot{\Theta} - 2\omega\dot{X} + \left(\omega_6^2 - \omega^2 - v_0^2 \, T\right)\cdot\Theta - u_0 \, v_0 \, T\cdot H = 0 \qquad \quad \\
\ddot{X} + 2\omega\,\dot{\Theta} + \left(\omega_6^2 - \omega^2\right)\cdot X = 0,
\end{IEEEeqnarray} \normalsize
\newpage
\noindent where $H \equiv \left(\eta_i\right)$, $\Theta \equiv \left(\theta_i\right)$, $X \equiv \left(\chi_i\right)$ and $Q \equiv \left(Q_{ik}\right)$, $T \equiv \left(T_{ik}\right)$.
Setting again
\begin{IEEEeqnarray}{c}
\left[\begin{array}{c} H \\ \Theta \\ X \end{array}\right] = e^{i\lambda t} \left[\begin{array}{c} \boldsymbol{\xi}_1 \\ \boldsymbol{\xi}_2 \\ \boldsymbol{\xi}_3 \end{array}\right], \ \label{Eigenvalues2}
\end{IEEEeqnarray}
we are led to the following system:
\small \begin{IEEEeqnarray}{l}
\left(u_0^2\,T + u_0 Q\right)\cdot\boldsymbol{\xi}_1 + u_0 v_0\,T\cdot\boldsymbol{\xi}_2 = \left(\omega_3^2 - \lambda^2\right)\boldsymbol{\xi}_1 \qquad \label{AngularEigenvalueProblem3a} \\
v_0^2\,T\,\boldsymbol{\xi}_2 = \left(\omega_6^2 - \omega^2 - \lambda^2\right)\boldsymbol{\xi}_2 - 2i\lambda\,\omega\,\boldsymbol{\xi}_3 - u_0 v_0\,T\,\boldsymbol{\xi}_1 \qquad \label{AngularEigenvalueProblem3b} \\
2i\lambda\,\omega\,\boldsymbol{\xi}_2 = \left(\lambda^2 - \omega_6^2 + \omega^2\right)\boldsymbol{\xi}_3. \qquad \label{AngularEigenvalueProblem3c}
\end{IEEEeqnarray} \normalsize
\indent The next step is to express the matrices $T$, $Q$ and $I$ in terms of the projection operators $P$, $R_{\pm}$ in \eqref{AngularEigenvalueProblem1a}--\eqref{AngularEigenvalueProblem2c}. Equations \eqref{AngularEigenvalueProblem3a}--\eqref{AngularEigenvalueProblem3c} become:
\begin{IEEEeqnarray}{c}
\left(\mathbb{A}_P \otimes P + \mathbb{A}_+ \otimes R_+ + \mathbb{A}_- \otimes R_-\right) \cdot \left[\begin{array}{c} \boldsymbol{\xi}_1 \\ \boldsymbol{\xi}_2 \\ \boldsymbol{\xi}_3 \end{array}\right] = 0 \qquad \quad \label{AngularEigenvalueProblem4a}
\end{IEEEeqnarray}
where
\begin{widetext}
\small\begin{IEEEeqnarray}{l}
\mathbb{A}_P = \left(\begin{array}{ccc}
\lambda^2 + s\left(u_0^2 - \frac{u_0}{2} + \frac{1}{18}\right) & s \, u_0 v_0 & 0 \\
s \, u_0 v_0 & \lambda^2 - s \, u_0^2 & 2 i \lambda \omega \\
0 & -2 i \lambda \omega & \lambda^2 - \frac{s}{2}\left(u_0 - \frac{1}{9}\right)
\end{array}\right), \qquad s \equiv j(j+1) - 2 \\[12pt]
\mathbb{A}_+ = \left(\begin{array}{ccc}
\lambda^2 + 2 j u_0^2 - j(j + 3) \frac{u_0}{2} + \frac{s}{18} & 2 j u_0 v_0 & 0 \\
2 j u_0 v_0 & \lambda^2 - 2 j u_0^2 - \frac{1}{2}\left(j^2 - j -2\right)\left(u_0 - \frac{1}{9}\right) & 2 i \lambda \omega \\
0 & -2 i \lambda \omega & \lambda^2 - \frac{s}{2}\left(u_0 - \frac{1}{9}\right)\end{array}\right) \\[12pt]
\mathbb{A}_- = \left(\begin{array}{ccc}
\lambda^2 - 2(j + 1)u_0^2 - [j(j-1) - 2]\frac{u_0}{2} + \frac{s}{18} & -2(j+ 1) u_0 v_0 & 0 \\
-2(j+ 1)u_0 v_0 & \lambda^2 + 2(j + 1)u_0^2 - \frac{1}{2}j(j+ 3)\left(u_0 - \frac{1}{9}\right) & 2 i \lambda \omega \\
0 & -2 i \lambda \omega & \lambda^2 - \frac{s}{2}\left(u_0 - \frac{1}{9}\right)\end{array}\right). \qquad
\end{IEEEeqnarray}
\end{widetext} \normalsize
\eqref{AngularEigenvalueProblem4a} gives rise to three different eigenvalue problems in the subspaces $P$ and $R_{\pm}$:
\begin{IEEEeqnarray}{c}
(\mathbb{A}_P \otimes P) \left[\begin{array}{c} \boldsymbol{\xi}_1 \\ \boldsymbol{\xi}_2 \\ \boldsymbol{\xi}_3 \end{array}\right]_P = (\mathbb{A}_{\pm} \otimes R_{\pm}) \left[\begin{array}{c} \boldsymbol{\xi}_1 \\ \boldsymbol{\xi}_2 \\ \boldsymbol{\xi}_3 \end{array}\right]_{\pm} = 0. \qquad \quad \label{AngularEigenvalueProblem4b}
\end{IEEEeqnarray}
Because the determinants of the projectors $R$ and $P_{\pm}$ are all equal to one, we can use the following property of determinants
\begin{IEEEeqnarray}{c}
\det\left(A \otimes B\right) = \left(\det A\right)^{\dim B}\left(\det B\right)^{\dim A}, \qquad
\end{IEEEeqnarray}
to transform \eqref{AngularEigenvalueProblem4b} to an eigenvalue problem for each of the three matrices $\mathbb{A}_P$, $\mathbb{A}_{\pm}$. The corresponding degeneracies will again be given by $d_P = 2j+1$, $d_+ = 2j+3$, $d_- = 2j-1$. \\[6pt]
\indent Notice that $\mathbb{A}_+$ and $\mathbb{A}_-$ are dual to each other under the involution $j \mapsto - j - 1$ while $\mathbb{A}_P$ is self-dual. We obtain three rather unusual eigenvalue problems, where the ($18(2j+1)$ in total) eigenvalues $\lambda$ also appear in the off-diagonal parts of $\mathbb{A}_P$, $\mathbb{A}_{\pm}$. Such eigenvalue equations are known to arise from characteristic polynomials with matrix coefficients \cite{Schwarz00a, CerchiaiZumino00}. \\[6pt]
\indent It turns out that one of the eigenvalues of $\mathbb{A}_P$ always vanishes while the other two can be computed exactly:
\begin{widetext}
\begin{IEEEeqnarray}{lll}
\lambda_{P}^2 = \frac{1}{2}\left(j^2 + j + 2\right)u_0 - \frac{1}{18}\left(1 + j\left(j + 1\right) \pm 3\sqrt{144\left(j^2 + j - 2\right)u_0^3 - 12\left(j^2 + j - 14\right)u_0^2 - 24u_0 + 1}\right). \qquad \quad \label{LyapunovExponents1}
\end{IEEEeqnarray}
\end{widetext}
Apart from the vanishing of the eigenvalue $\lambda_{P(-)}$ when $j = 1$, $\lambda_{P(\pm)}^2$ are always positive (and thus stable) for any $j$ in the interval $1/6 < u_0 < 1/3$. \\[6pt]
\indent The expressions for the three eigenvalues of $\mathbb{A}_+$ (as well as those of $\mathbb{A}_-$ obtained by the substitution $j \mapsto -j-1$) are rather complicated functions of $j$ and $u_0$. Writing the characteristic equation of $\mathbb{A}_{\pm}$ as
\begin{IEEEeqnarray}{lll}
x^3 + \alpha_{\pm} x^2 + \beta_{\pm} x + \gamma_{\pm} = 0, \qquad x \equiv \lambda^2, \qquad \label{CharacteristicEquation}
\end{IEEEeqnarray}
where $\alpha_{\pm}, \ \beta_{\pm}, \ \gamma_{\pm}$ are real polynomial functions of $u_0$ and $j$, it is easy to see that for $j \geq 3$, $\alpha_{\pm} < 0$, $\beta_{\pm} > 0$ and $\gamma_{\pm} < 0$ inside the interval \eqref{ExtremalBounds}. The Descartes rule of signs then implies that \eqref{CharacteristicEquation} has either three real positive roots, or one real positive root and two complex ones for any $j \geq 3$. Because the discriminant of \eqref{CharacteristicEquation},
\begin{IEEEeqnarray}{lll}
\Delta = 18\alpha \beta \gamma - 4\alpha^3 \gamma + \alpha^2 \beta^2 - 4\beta^3 - 27\gamma^2 \qquad
\end{IEEEeqnarray}
is positive for every value of $j \geq 1$ inside the interval \eqref{ExtremalBounds}, the equation \eqref{CharacteristicEquation} can only have real roots, which turn out to be positive for $j \geq 3$ (by the rule of signs). This again implies the (multipole) stability of the corresponding extrema. Alternatively, we can reach the same conclusion from the monotonicity of each term of the characteristic polynomial \eqref{CharacteristicEquation} and the relative positions of its maxima and minima on the real axis. \\[6pt]
\indent For $j = 1$ one of the eigenvalues of $\mathbb{A}_{\pm}$ vanishes, whereas the other two coincide with those obtained from the analysis of the radial perturbations in our previous paper \cite{AxenidesFloratosLinardopoulos17a}:\footnote{The overall minus sign difference with our radial perturbation analysis of \cite{AxenidesFloratosLinardopoulos17a} appears because in \eqref{Eigenvalues2} we have used $e^{i\lambda t}$ rather than $e^{\lambda t}$.}
\small\begin{IEEEeqnarray}{l}
\lambda_{+}^2 = \frac{5u_0}{2} - \frac{1}{9} \pm \sqrt{\frac{1}{9^2} - \frac{u_0}{9} - \frac{5u_0^2}{12} + 4 u_0^3}, \ \label{LyapunovExponents2} \\[6pt]
\lambda_{-}^2 = \frac{5u_0}{2} - \frac{5}{18} \pm \sqrt{\frac{5^2}{18^2} - \frac{35u_0}{18} + \frac{163u_0^2}{12} - 20u_0^3}. \qquad \quad \label{LyapunovExponents3}
\end{IEEEeqnarray}\normalsize
These are all positive and thus stable in the interval \eqref{ExtremalBounds}, except $\lambda_{-(-)}^2$ which is positive/stable only for $u_{\text{crit}} < u_0 < 1/3$ . \\[6pt]
\indent For $j = 2$, the roots of the cubic equation \eqref{CharacteristicEquation} are (very complicated but) explicitly known so that we may plot the squares of the eigenvalues of $\mathbb{A}_{\pm}$ in terms of $u_0$. It is obvious from figures \ref{Graph:Eigenvalues1}--\ref{Graph:Eigenvalues2} that all the eigenvalues of $\mathbb{A}_+$ are positive for $j=2$ (implying the stability of the corresponding membrane modes) inside the interval \eqref{ExtremalBounds}. On the other hand, one eigenvalue of $\mathbb{A}_-$ takes negative values in the interval
\begin{IEEEeqnarray}{lll}
\frac{1}{6} \leq u_0 \leq 0.207245 < u_{\text{crit}},
\end{IEEEeqnarray}
when $j=2$. It therefore corresponds to an unstable mode of the configuration. Notice that the unstable region becomes smaller as we go from $j=1$ to $j=2$ and completely disappears for $j \geq 3$. This of course holds only within the linearized approximation. \\[6pt]
\indent As in the case of the $SO(3)$ sector, the validity of the linearized Gauss constraint \eqref{GaussLaw3} follows from its validity at $t=0$. By using the rotated variables $\theta$ and $\chi$ (defined in \eqref{Rotation1}--\eqref{Rotation2}) we may eliminate the time dependence from the constraint equation, getting:
\begin{IEEEeqnarray}{ll}
\sum_{m}\sum_{i=1}^3 \left(J_i\right)_{m'm}\left(u_0\dot{\eta}_{i}^{jm}\left(0\right) + v_0\dot{\theta}_{i}^{jm}\left(0\right)\right) = 0, \qquad \quad \label{CoplanarityConstraint2}
\end{IEEEeqnarray}
which is similar to \eqref{CoplanarityConstraint1}. The system \eqref{CoplanarityConstraint2} always has a solution that can be expressed as a linear combination of the eigenvectors $\boldsymbol{\xi}_1$, $\boldsymbol{\xi}_2$ and $\boldsymbol{\xi}_3$. Therefore the Gauss constraint \eqref{GaussLaw3} is satisfied at $t = 0$ and at all times $t$. \\[6pt]
\indent In sum we find that the spectrum of the multipole perturbations corresponding to the ansatz \eqref{Ansatz2}--\eqref{Ansatz5} always possesses at least one vanishing eigenvalue for every value of $j$. A second vanishing eigenvalue (namely the eigenvalue $\lambda_{P(-)}$ of $\mathbb{A}_P$ in \eqref{LyapunovExponents1}) emerges for $j = 1$. For $j =1,2$ the spectrum contains exactly one unstable direction in the intervals $1/6 < u_0 < u_{\text{crit}}$ and $1/6 < u_0 < 0.207245$, in the form of purely imaginary eigenvalues of $\mathbb{A}_-$ (for $j = 1$ this is the eigenvalue $\lambda_{-(-)}$ in \eqref{LyapunovExponents3}). For $j \geq 3$ (and with the exception of one vanishing eigenvalue) all of the eigenvalues of $\mathbb{A}_P$ and $\mathbb{A}_{\pm}$ are purely real and thus the system is stable. \\[6pt]
\indent These results are summarized in the following table which contains the sign of the square of each eigenvalue in the interval $1/6 < u_0 < 1/3$, for all values of the angular momentum $j$.
\small\vspace{-.4cm}\begin{center}
\begin{eqnarray}
\begin{array}{|c|c|c|c|}
\hline &&& \\
\text{eigenvalues} & j=1 & j=2 & j \geq 3 \\[6pt]
\hline &&& \\
\lambda_P^2 & 0,0,+ & 0,+,+ & 0,+,+ \\[12pt]
\lambda_+^2 & 0,+,+ & +,+,+ & +,+,+ \\[12pt]
\lambda_-^2 & 0,+,\{0,\pm\} & +,+,\{0,\pm\} & +,+,+ \\[3pt]
& \footnotesize{\Big(\begin{array}{c} \text{positive for} \\ u_0 > u_{\text{crit}} \end{array}\Big)} & \footnotesize{\Big(\begin{array}{c} \text{positive for} \\ u_0 > 0.207245 \end{array}\Big)} & \\[6pt]
\hline
\end{array} \nonumber
\end{eqnarray}
\end{center}\normalsize
%
%
\section[Summary \& Conclusions]{Summary \& Conclusions\label{Section:Discussion}}
\noindent In this paper we have studied a class of classical solutions of the BMN matrix model that are described at large-$N$ by spinning ellipsoidal membranes in either the $SO(3)$ or the $SO(3)\times SO(6)$ subsector of the 11-dimensional plane-wave background. In order to examine the stability of these specific classical solutions, we have perturbed their ellipsoidal shape by means of small multipole perturbations that are defined in terms of the $j,m$ spherical harmonic ($j = 1,2,\ldots$, $m =-j,\ldots j$). We then went on to study the time evolution of the spherical harmonic coefficients as the system relaxes. \\[6pt]
\indent Working at the level of linearized perturbation theory, we have found that, for each value of $j$, all modes with different values of $m$ do couple between themselves. On the other hand there's no coupling between modes with different $j$'s. For each value of $j$ we have thus obtained a linear system for the perturbation coefficients from which we can determine the corresponding Lyapunov spectrum by diagonalizing the associated fluctuation matrix. In both the $SO(3)$ and $SO(3)\times SO(6)$ sectors, we have found that only the $j=1,2$ modes (corresponding to dipole and quadrupole deformations respectively) can possibly be unstable when their $SO(3)$ radius is in the interval $1/6 \leq u_0 \leq u_{\text{crit}}$. \\[6pt]
\indent We have also proven that all the perturbations (i.e.\ for $j=1,2,\ldots$) are stable in the interval $0.207245 < u_0 \leq 1/3$ and, for $j>2$, they are stable in the entire interval, $1/6 \leq u_0 \leq 1/3$. \\[6pt]
\indent By examining higher orders in perturbation theory beyond the linear level (in the interval $1/6 \leq u_0 \leq u_{\text{crit}}$) we expect to obtain a cascade of instabilities that originates from the $j = 1,2$ sectors and propagates towards the higher multipoles. This is due to the fact that the various (constant $j$) multipoles at a given order in perturbation theory couple to all the $j$’s of the previous orders through an effective forcing term that arises in the corresponding fluctuation equation. E.g.\ the lowest order instabilities (at $j = 1,2$) are coupled to all the modes (having different $j$'s) of the first order. \\[6pt]
\indent Thus the instabilities propagate towards the higher modes in a way that depends on the algebra of area-preserving diffeomorphisms of the perturbed configuration. Eventually, we expect to find chaos to all orders of perturbation theory. The physical interpretation of these instabilities at the quantum level can probably be related to the spontaneous emission of higher spin states. \\[6pt]
\indent It is obvious that more numerical and analytical studies (by using methods of classical chaos such as KAM theory, etc.)\ are needed in order to make progress in understanding the full spectrum of Lyapunov exponents in the infinite-dimensional phase space of the multipole modes $x^{jm}\left(\tau\right)$ and $y^{jm}\left(\tau\right)$. Some numerical results in the context of the BMN and BFSS matrix models have recently appeared in \cite{AsanoKawaiYoshida15, Gur-AriHanadaShenker15}. See also \cite{CotlerCotlerGur-AriHanadaPolchinskiSaadShenkerStanfordStreicherTezuka16}.
\section[Acknowledgements]{Acknowledgements}
\noindent The authors would like to thank Jens Eggers, Jens Hoppe, Yannis Kominis, Stam Nicolis, Georgios Pastras and Euaggelos Zotos for illuminating discussions. We are especially thankful to Konstantinos Efstathiou and Christos Efthymiopoulos for explaining their work to us. E.F.\ and G.L.\ kindly acknowledge Gian Giudice and the CERN Theory Group for instructive discussions and generous support. G.L.\ is grateful to Jens Hoppe and the KTH Royal Institute of Technology for hospitality and support. 
\appendix
\bibliography{HEP_Bibliography,Math_Bibliography}
\end{document}